\documentstyle[11pt,twoside]{article}
\def\Red{}
\def\Black{}
\def\Blue{}

\newcommand{\NP}[3]{{\em Nucl. Phys. \bf #1}  (#2) #3}
\newcommand{\PRL}[3]{{\em Phys. Rev. Lett. \bf #1} (#2) #3}
\newcommand{\PL}[3]{{\em Phys. Lett. \bf #1} (#2) #3}
\newcommand{\PR}[3]{{\em Phys. Rev. \bf #1} (#2) #3}

\makeatletter
\newcounter{alphaequation}[equation]
\def\thealphaequation{\theequation\hbox to
0.6em{\hfil\alph{alphaequation}\hfil}}
\def\eqnsystem#1{
\def\@eqnnum{{\rm (\thealphaequation)}}
\def\@@eqncr{\let\@tempa\relax \ifcase\@eqcnt \def\@tempa{& & &} \or
  \def\@tempa{& &}\or \def\@tempa{&}\fi\@tempa
  \if@eqnsw\@eqnnum\refstepcounter{alphaequation}\fi
\global\@eqnswtrue\global\@eqcnt=0\cr}
\refstepcounter{equation} \let\@currentlabel\theequation \def\@tempb{#1}
\ifx\@tempb\empty\else\label{#1}\fi
\refstepcounter{alphaequation}
\let\@currentlabel\thealphaequation
\global\@eqnswtrue\global\@eqcnt=0 \tabskip\@centering\let\\=\@eqncr
$$\halign to \displaywidth\bgroup \@eqnsel\hskip\@centering
$\displaystyle\tabskip\z@{##}$&\global\@eqcnt\@ne
\hskip2\arraycolsep\hfil${##}$\hfil& \global\@eqcnt\tw@\hskip2\arraycolsep
$\displaystyle\tabskip\z@{##}$\hfil
\tabskip\@centering&\llap{##}\tabskip\z@\cr}
\def\endeqnsystem{\@@eqncr\egroup$$\global\@ignoretrue} \makeatother
\makeatletter
\newcommand{\newsection}%
{\@startsection{section}{1}{0em}{-\baselineskip}%
{-0em}{\large\bf}}
\makeatother

\newcommand{\Dslash}{D\hspace{-7pt}/}
\newcommand{\mrm}[1]{{\rm #1}}
\newcommand{\eqref}[1]{~{\rm (\ref{#1})}}
\newcommand{\Ord}{{\cal O}}

\newcommand{\GeV}{\,\mrm{GeV}}
\newcommand{\MeV}{\,\mrm{MeV}}

\newcommand{\ecm}{\,e\cdot \mrm{cm}}

\def\Tr{\mathop{\rm Tr}}

\newcommand{\gs}{g_{a\!\hskip1pt N\!N}}

\newcommand{\lchi}{\Lambda_{\chi}}
\newcommand{\Oc}{{\cal O}^{\chi}}
\newcommand{\Onc}{{\cal O}^{{\rm n}\chi}}
\newcommand{\circa}[1]{\,\raise.3ex
\hbox{$#1$\kern-.75em\lower1ex\hbox{$\sim$}}\,}

\begin{document}\thispagestyle{empty}
May 1996\hfill
\vbox{\hbox{\bf IFUP--TH 28/96}\hbox{\bf FT--UAM 96/21}
      \hbox{\bf hep-ph/9605368}}\\[5mm]

\centerline{\LARGE\bf\Red On axion-mediated macroscopic forces again}
\vspace{0.5cm}
\bigskip\bigskip\Black
\centerline{\large\bf  Riccardo Barbieri, Andrea Romanino}
\vspace{4mm}
\centerline{\large\em Dipartimento di
Fisica, Universit\`a di Pisa \rm and}
\centerline{\large\em INFN, Sezione di Pisa, I-56126 Pisa, Italy}
\bigskip
\centerline{\large\bf and Alessandro Strumia}\vspace{4mm}
\centerline{\large\em Departamento de F\'{\i}sica Te\'orica,
             m\'odulo \rm C--XI}
\centerline{\large\em Universidad Aut\'onoma de Madrid, 28049,
             Madrid, Espa\~na}
\centerline{\large\em {\rm and} INFN, Sezione di Pisa, I-56126 Pisa, Italia}
\bigskip\bigskip\bigskip\Blue \centerline{\large\bf Abstract}
\begin{quote}\large\indent
In a supersymmetric unified theory or in a generic model where a large
neutron electric dipole moment $d_N$ is expected, close to the present
bound, we estimate the relation
$\gs=10^{-21\pm1}(d_N/10^{-25}\ecm)(10^{10}\GeV/f_a)$ between $d_N$
itself, the scalar coupling $\gs$ to nucleons of the axion, assumed to
exist, and the breaking scale, $f_a$, of the Peccei-Quinn symmetry.
Newly developing techniques to search for sub-cm macroscopic forces
might reveal a signal due to axion exchange at least in a favorable
range of $f_a$.
\end{quote}\Black

\setcounter{page}{0}
\newpage
\newsection{}
Electric Dipole Moments (EDMs) of the electron and the
neutron at the border of the present limits are expected in
supersymmetric unified theories with supersymmetry breaking transmitted
by supergravity
couplings~\cite{dimopoulos:95a}.
Such EDMs are generated from CKM-like
phases via one loop diagrams involving sfermion and gaugino-higgsino
exchanges at the weak scale.
In these models, however, as in most other cases where a sizeable
one-loop quark-EDM occurs~\cite{barr:92a}, similar diagrams give also
rise to a strong 
CP violating angle, $\theta_{\rm QCD}$, which is
{\em too large\/} if not counteracted
by an appropriately tuned initial condition.
Therefore, especially in these models, a
Peccei-Quinn~\cite{peccei:77a} (PQ) solution of
the strong CP problem is called for,
leading to a so-called 
``invisible'' axion~\cite{dine:81a}.

Unfortunately, such a solution of the strong CP problem is as elegant
as it is experimentally elusive.
Nevertheless, rightly so, a number of serious experimental proposal
for axion detection have been made. 
Among them, the search for a CP-violating macroscopic force mediated
by axion exchange~\cite{moody:84a} is the possibility that we want to
reconsider in this 
letter.

\newsection{}
Crucial parameters to this effect are the mass $m_a$ and the
{\em scalar\/}
coupling $\gs$ of the axion to the nucleons.
Both $m_a$ and $\gs$ are inversely proportional to the PQ symmetry
breaking scale, conventionally called $f_a$, which is constrained to lie
in the range $10^7\GeV\circa{<} f_a\circa{<}
10^{12}\GeV$~\cite{fukugita:82a}.
In terms of the quark masses, $m_u$, $m_d$, and of the pion mass and
decay constant, $m_\pi$ and $f_\pi$, it is
\begin{equation}\label{eq:ma}
m_a=\frac{m_\pi f_\pi}{f_a}\frac{\sqrt{m_u m_d}}{m_u+m_d}=
\frac{1}{0.02\,{\rm cm}}\frac{10^{10}\GeV}{f_a}.
\end{equation}
Furthermore, the required weak CP violation leads to a residual
dynamically determined $\theta_{\rm QCD}\neq 0$, which, in turn, induces an
axion-nucleon scalar coupling~\cite{moody:84a}
(disregarding the relatively small but
phenomenologically potentially important difference between the
axion-proton and the axion-neutron couplings) 
\begin{equation}\label{eq:gtheta}
\gs[\theta_{\rm QCD}]=
\frac{\theta_{\rm QCD}}{f_a}\frac{m_u m_d}{m_u+m_d} \langle
N|\bar{u}u+\bar{d}d|N\rangle
\approx 3\cdot 10^{-12}\, \theta_{\rm QCD}\,\frac{10^{10}\GeV}{f_a}
\end{equation}
In view of the limit set by the null results of the measurements of
the neutron EDM so far~\cite{altarev:92a},
$\theta_{\rm QCD}\circa{<} 10^{-9}$ \cite{baluni:79a},
the Yukawa-type
interaction induced by one-axion exchange is therefore bound to be
small, at about the level of gravity or lower.
Maybe not so small, however, to escape detection in experiments
proposed~\cite{price:88a} or conceivable~\cite{moody:84a}
to search for new
sub-cm forces. 
The potentiality of axion searches by looking for axion-mediated
macroscopic forces has been already emphasized in ref.~\cite{thomas}.

All this makes it interesting to ask, in supersymmetric unified theories, at
what level $\gs[\theta_{\rm QCD}]$ actually
sets in (what is $\theta_{\rm QCD}$?), or,
more importantly, what 
is the value of $\gs$ at all, including any possible effect from other
CP violating operators. To our knowledge these questions have been
addressed and satisfactorily answered~\cite{georgi:86b}
only in the case of the Standard
Model, reaching a pretty negative conclusion: in the SM $\gs$ is too
small to be of any interest. 
One should not forget, however, that CP violation in the electroweak
sector of the SM is screened enough that, even in absence of an axion,
the radiative contributions to the $\theta_{\rm QCD}$ parameter are also
negligibly small~\cite{ellis:79a}.

\newsection{}
Of relevance to the question under consideration is the effective
lagrangian just above the chiral symmetry breaking scale, $\lchi$,
including the axion interactions and the flavour-conserving
CP-violating operators.
As we shall see, it is useful to consider at the same time the axion
coupling and the neutron EDM, since the relation between the two
quantities is largely model independent.

Following ref.s~\cite{georgi:86b,georgi:86a}, we consider a non linear
realization 
of the PQ 
symmetry where the axion field $a$ transforms as
\[
a\rightarrow a+\mrm{cte},
\]
whereas all the matter fields remain invariant.
In this basis the axion would have no non-derivative coupling at all,
if it were not for the anomalous term
\begin{equation}\label{eq:anomalous}
-\frac{a}{f_a}\frac{\alpha_{\rm s}}{8\pi}\,G^{\mu\nu}\tilde{G}_{\mu\nu}.
\end{equation}
This term too can be eliminated by a chiral rotation acting on the
quark fields $q=(u,d)^T$
\begin{equation}\label{eq:rotation}
q\rightarrow \exp(-iQ_A\gamma_5 \frac{a}{f_a})\,q
\end{equation}
at the price of introducing axion dependence in the chirality breaking
quark operators. In terms of the quark mass matrix $m_q$, the matrix
$Q_A$ is
\begin{equation}\label{eq:generatore}
Q_A=\frac{1}{2}\frac{m_q^{-1}}{\Tr m_q^{-1}},
\end{equation}
chosen to eliminate mass mixing between the axion and the pseudoscalar
mesons.
In the effective lagrangian it is therefore useful to distinguish,
among the CP violating operators, those ones that respect chiral
symmetry, generically denoted by $\Oc$, from those that break chiral
symmetry, $\Onc$.
After elimination of the anomaly term\eqref{eq:anomalous} by the
chiral rotation\eqref{eq:rotation}, the
relevant axion dependence resides in the mass term
\begin{equation}\label{eq:mass}
\overline{q_L}~e^{-iQ_A a/f_a}~M~e^{-iQ_A a/f_a}~q_R
\end{equation}
and in $\Onc$ only.

Examples of $\Oc$ are the Weinberg 3-gluon operator~\cite{weinberg:89a}
\begin{eqnsystem}{sys:1}
&G^{\mu}_{\nu}G^{\nu}_{\rho}\tilde{G}^{\rho}_{\mu},&\label{eq:weinberg}\\
\noalign{\hbox{six-quark operators like}} \nonumber \\[-2mm]
&[\overline{u_L}\gamma^{\mu}d_L\cdot\overline{d_L}\gamma_{\mu}]
\Dslash~\label{eq:sixquark}
[\gamma^{\nu}s_L\cdot\overline{s_L}\gamma_{\nu}u_L],&\\[2mm]
\noalign{\hbox{or the interplay between two flavour violating operators}}
\nonumber\\
&(\overline{s_L}\gamma_{\mu}d_L)(\overline{u}\gamma_{\mu}u)
\quad\hbox{and}\quad \label{eq:interplay}
(\overline{s_L}\gamma_{\mu}d_L)(\overline{d}\gamma_{\mu}d).&
\end{eqnsystem}
The prototype example of $\Onc$ is, on the other hand, the
ChromoElectric Dipole Moment (CEDM) operator
\begin{equation}\label{chromo}d^{\rm QCD}_q\times
\frac{1}{2}(\bar{q}\sigma_{\mu\nu}\gamma_5 q) G^{\mu\nu}.
\end{equation}

To obtain the axion-nucleon scalar coupling and the neutron EDM one
has to cross the chiral symmetry breaking scale $\lchi\approx 1\GeV$
and go to the confinement scale, just above $\Lambda_{\mrm{QCD}}$.
This we do, as in ref.~\cite{georgi:84a}, by use of Naive Dimensional
Analysis (NDA). 
This technique is appropriate to the general discussion that we want
to make and is not too inaccurate, given our presently limited
understanding of low energy QCD. The crucial notion of NDA is that
the reduced coupling $\bar{g}$ appearing in front of an operator
$\Ord$, that one seeks to calculate in the effective
hadronic theory, is given by the product of the reduced couplings of
the operators that produce $\Ord$ in the effective
lagrangian involving quarks and gluons.
For an operator with dimensionful coupling $g$, of dimension $d$ in
mass and involving $n$ fields, the dimensionless reduced coupling
$\bar{g}$ is
\begin{equation}\label{eq:reduced}
\bar{g}\equiv g\cdot (4\pi)^{2-n}\lchi^{d-4}.
\end{equation}

\begin{figure}[t]\setlength{\unitlength}{1cm}
\begin{center}
\begin{picture}(15,9)
\put(-1.7,-0.5){\includegraphics{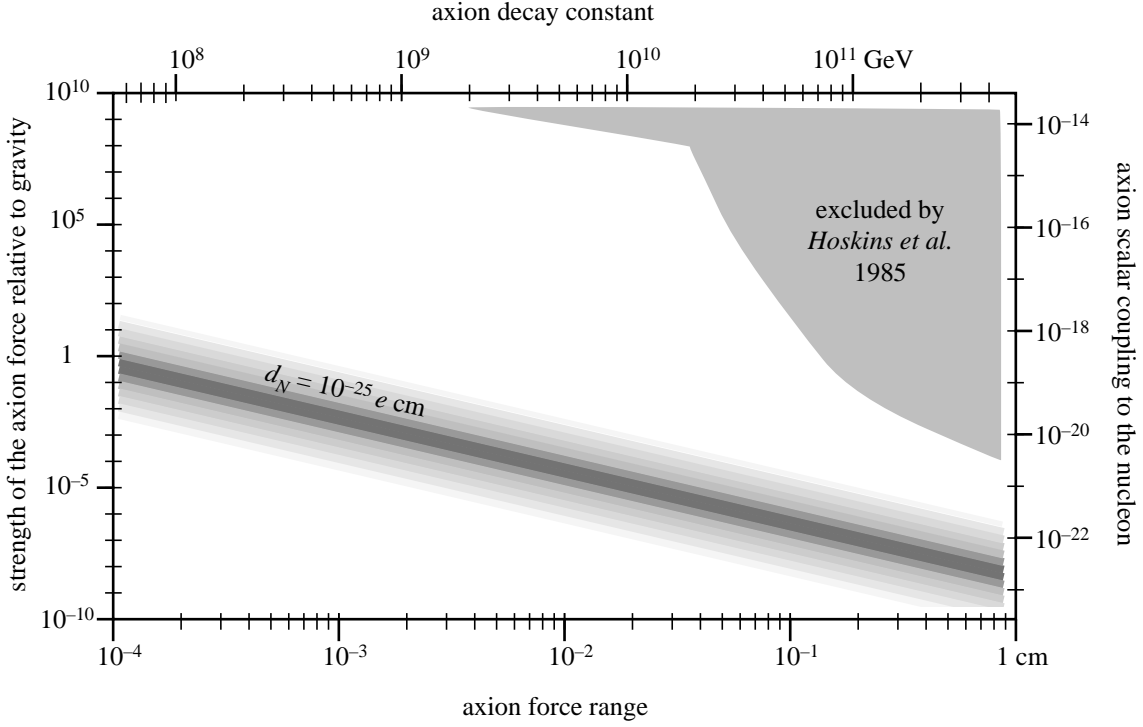}}
\end{picture}
\caption[a]{\em estimates of the axion force strength relative to gravity for
$d_N=10^{-25}\ecm$. Also shown is the presently excluded region
.\label{fig:assiun}}
\end{center}\end{figure}

\newsection{}
As mentioned, we consider at the same time the axion-nucleon scalar
coupling, $\gs$, and the neutron EDM, $d_N$.
Notice that they not only both violate CP but also have the same
chiral properties.

As source of CP violation, let us take first the quark CEDM $d^{\rm QCD}_q$.
The CEDM operator carries axion dependence, since it breaks chiral
symmetry; it has in fact the same chiral properties of $\gs$ and $d_N$
themselves.

By means of NDA, it is immediate to get the contributions to $d_N$ and
$\gs$ induced by $d^{\rm QCD}_q$. It is 
\begin{eqnsystem}{sys:xx}
d_N[d^{\rm QCD}_q]&\approx&\makebox(0,0)[b]{$\quad e$}
\phantom{\frac{\lchi^2}{f_a}}
\frac{g_{\rm s}}{(4\pi)^2} \langle d^{\rm QCD}_q\rangle,
\label{eq:contributions}\\
\gs[d^{\rm QCD}_q]&\approx&\frac{\lchi^2}{f_a}
\frac{g_{\rm s}}{(4\pi)^2}
\langle 2Q_A d^{\rm QCD}_q\rangle, 
\end{eqnsystem}
where $\langle\cdots\rangle$ denotes a weighted sum, with coefficients of
order unity, over the up and down quarks.
From\eqref{eq:contributions}, since
$\langle d^{\rm QCD}_q\rangle\approx\langle2Q_A d^{\rm QCD}_q\rangle$, we
have 
\begin{equation}\label{eq:gsdnd}
\gs[d^{\rm QCD}_q]\approx\frac{d_N[d^{\rm QCD}_q]}{e}\frac{\lchi^2}{f_a}.
\end{equation}
It should be clear, however, that a relation like\eqref{eq:gsdnd}
holds, within the limits of NDA, for any operator, or combination of
operators, of the type $\Onc$, involving quarks and gluons only,
\begin{equation}\label{eq:chib}
\gs[\Onc]\approx\frac{d_N[\Onc]}{e}\frac{\lchi^2}{f_a}.
\end{equation}
A different relation holds between $\gs$ and $d_N$ generated by the
standard quark EDMs $d_q$, since
\begin{eqnsystem}{sys:2}
d_N[d_q]&\approx& \langle d_q\rangle
\label{eq:contributions2}\\
\gs[d_q]&\approx& 
\frac{e}{(4\pi)^2}\frac{\lchi^2}{f_a}\langle 2Q_A d_q\rangle.
\end{eqnsystem}
In the models of interest, however, $d_N[d^{\rm QCD}_q]\circa{>} d_N[d_q]$, so
that $\gs[d_q]<\gs[d^{\rm QCD}_q]$. 
Eq.\eqref{eq:chib} remains therefore appropriate even with the
inclusion in $\Onc$ of the quark EDMs.

Let us now consider $\gs$ and $d_N$ generated by CP-violating chirally
invariant operators $\Oc$.
In this case an asymmetry occurs between $\gs$ and $d_N$.
Although, to generate both $\gs$ and $d_N$, $\Oc$ must be supplemented
by a chirality breaking operator, the most economic way for $d_N$ is
through the so called ``soft'' quark mass
\begin{equation}\label{soft}
m_{\rm soft}\approx g_{\rm s}^2
\frac{\langle\bar{q}q\rangle}{p^2}\approx 350 \MeV
\end{equation}
(becoming actually soft only at momenta $p\gg\lchi$), whereas $\gs$
comes through the ``current'' mass term\eqref{eq:mass} in order to
introduce also the required $a$-dependence.
Therefore eq.\eqref{eq:chib} is corrected by a relative factor
$\langle 2 Q_A M\rangle/m_{\rm soft}$ or 
\begin{equation}\label{eq:chip}
\gs[\Oc]\approx\frac{m_u m_d}{m_{\rm soft}(m_u+m_d)}
\frac{d_N[\Oc]}{e}\frac{\lchi^2}{f_a}.
\end{equation}
Eq.s\eqref{eq:chib} and\eqref{eq:chip} represent our estimates for
the relation between $\gs$ and $d_N$ in a generic model, which can of
course be summarized as
\begin{equation}\label{eq:summ}
\gs\approx\frac{\lchi^2}{ef_a}\left\{d_N[\Onc] +\frac{m_u
m_d}{m_{\rm soft}(m_u+m_d)}d_N[\Oc]\right\}.
\end{equation}
The SM is a prototype of models where $d_N(\Oc)$
dominates $d_N$ (which is, mostly for the same reason, rather
small)~\cite{ellis:79a,shabalin:78a}. Consequently
$$
\gs^{\rm SM}\approx\frac{\lchi^2}{ef_a}\frac{m_u
m_d}{m_{\rm soft}(m_u+m_d)}d_N^{\mrm{SM}} \approx
10^{-30\pm1}\frac{d_N^{\mrm{SM}}}{10^{-32}\ecm}
\frac{10^{10}\GeV}{f_a},
$$
too small to be of any experimental interest~\cite{georgi:86b}.
On the other hand, for the Unified Supersymmetric Models
(USMs) or for a generic model where $d_N$ is dominated
by $d_N[\Onc]$ (and possibly large, because of this very reason)
\begin{equation}\label{eq:last}
\gs^{\mrm{USM}}\approx10^{-21\pm1}
\frac{d_N^{\mrm{USM}}}{10^{-25}\ecm}\frac{10^{10}\GeV}{f_a}.
\end{equation}
We have explicitly indicated the uncertainty that must be attributed
to our estimates, essentially due to the limited control of QCD in
the infrared regime.

Taking into account of the expectations for $d_N^{\mrm{USM}}$,
which saturate the present bound~\cite{dimopoulos:95a},
{\em the value of $\gs$
in eq.\eqref{eq:last} leads to a signal at the border of the
sensitivity of planned or conceived
experiments to search for macroscopic sub-{\rm cm} forces\/}, at least in a
favorable range of $f_a$~\cite{price:88a,moody:84a}.

For the dimensionless ratio between the strength of the axion induced
gravity-like force and gravity itself, one has
\begin{equation}\label{eq:aggiunta}
\frac{{\cal F}_{\rm axion}}{{\cal F}_{\rm gravity}} =
+ \frac{\gs^2/4\pi}{G_{\rm N}^{} m^2_N}
=10^{-5\pm 2}\,\left(\frac{d_N}{10^{-25}\ecm}\right)^{\!\!2}
\left(\frac{10^{10}\GeV}{f_a}\right)^{\!\!2},
\end{equation}
as represented in fig.~\ref{fig:assiun}.
Monopole-dipole effects might also be relevant~\cite{moody:84a}.
E\"otv\"os-type experiments, if possible in the sub-cm range, would of
course also be of great significance.
The importance of looking for such effects cannot be possibly
overestimated.
It is interesting to notice that the relevance of similar types of
experiments has also been recently emphasized in connection with the
moduli fields characteristic of superstring theories~\cite{dimo}.

\medskip

To conclude, we notice that the contribution to $\gs$ from the
$\theta_{\rm QCD}$-parameter, which started our discussion, does not alter in
any significant way the result in eq.\eqref{eq:summ}. Actually, the
very distinction between $\gs[\theta_{\rm QCD}]$ and the other contributions
to $\gs$ is not even unambiguously defined in the hadronic lagrangian
because $\theta_{\rm QCD}$ itself is not, unlike the case for the basic QCD
lagrangian, since several terms will generally have independent
phases, each of the same order.

\newsection*{Acknowledgements}~\\
We thank Savas Dimopoulos for a useful comment.

\end{document}